# PROMPT: A Pre-registered Randomized Protocol for Component-Level Evaluation of Clinical AI Prompts

**Short description:** PROMPT is a reusable protocol for randomized component-level evaluation of clinical AI prompts using matched controls, factorial prompt arms, locked stimuli, and blinded scoring

Authors: Bin Hu[1,2], Avneek Sandhu[2] and Shahryar Wasif[2]

Affiliations:

1. Division of Translational Neuroscience, Department of Clinical Neurosciences, Hotchkiss Brain Institute, University of Calgary, Calgary, AB T2N 4N1, Canada
2. Canadian Open Digital Health (OpenDH) program, University of Calgary, Calgary, Alberta, Canada T2N 4N1

To whom corresponding should be addressed:
Bin Hu MD. Ph.D.
Endowed Professor in Translational Neuroscience
Founder and Director, OpenDH program and laboratory of Translational AI
Department of Clinical Neurosciences, Hotchkiss Brain Institute, University of Calgary, Calgary, AB T2N 4N1, Canada
Email: hub@ucalgary.ca

## Abstract

### BACKGROUND

Prompt engineering shapes medical AI outcomes, but prompt components are rarely tested as clinical interventions. We developed PROMPT (Pre-registered Randomized Outcome Measurement for Prompt Testing), a protocol using pre-specification, randomization, matched controls, dismantling, and decision rules.

### METHODS

Two pre-registered demonstrations used Claude Sonnet 4.6. Experiment 1 used a synthetic tumbling-E orientation task: 630-trial main study, 480-trial dismantling study, and 1,050-trial 2×2 factorial extension. Experiment 2 used the same arms on 16 label-masked CBIS-DDSM mammographic crops in four orientations: 256 confirmatory trials and a 64-trial Arm E extension. Matched controls removed the active component while preserving framing, structure, and output format.

### RESULTS

PROMPT identified beneficial, inactive, harmful, and task-dependent effects. In Exp 1, the full prompt achieved 98.6% orientation accuracy; removing the decoding rule reduced accuracy to 50.1% (difference, +48.5 pp; 95% CI, +43.2 to +53.7; $P<0.001$). A rule-only arm matched the full prompt (maximum difference, 2.3 pp), identifying the decoding rule as the sole measurable active component. A prohibited-reasoning block assumed to improve safety was inactive, an effect missed by whole-prompt comparison. Scaffolding without the task-specific rule underperformed the vehicle prompt, showing prompt structure alone was harmful. Exp 1 revealed a canonical-RIGHT error phenotype in no-rule arms, consistent with a RIGHT-orientation prior. In Exp 2, the phenotype recurred on mammographic images, but the rule's benefit was attenuated and did not meet the threshold (+14.1 pp; bootstrap 95% CI, −3.1 to +29.7; post-hoc mixed-model 95% CI, +3.5 to +24.6).

### CONCLUSION

PROMPT revealed component effects missed by whole-prompt evaluations, identifying safety vulnerabilities and performance failures before clinical AI deployment.

## Introduction

Large language models are increasingly used in clinical documentation, triage, diagnostic assistance, decision support, and medical image interpretation[1,2,11,12]. In many applications, the foundation model is not retrained for each use case. Rather, model behavior is configured through prompts, scaffolding blocks, task rules, personas, and output schemas[1-3]. Prompts therefore function as part of the clinical AI intervention rather than as cosmetic implementation details[1-3].

Prompt sensitivity is well documented. Rephrasing, role assignment, chain-of-thought instructions, output constraints, and context position can change performance and failure modes. Yet despite their influence on model behavior, the efficacy and safety of prompts are rarely evaluated with the same prospective, randomized, and pre-specified methods expected of other clinical AI system components. A model may produce different outputs when asked to reason step-by-step,[3] when given a system-role persona, or when the same clinical question is rephrased,[4-7] and these structural differences can be clinically consequential[5,6].

In medical AI studies, prompts are often selected through informal iteration and evaluated as whole units.[5],[6] Standard A/B comparisons can show that one prompt performs better than another, but they usually cannot determine which component caused the effect, whether another component was redundant, or whether a removed component created a harmful interaction. A prompt element that appears neutral in aggregate may still be harmful for a specific disease subtype, image orientation, or clinical context. Recent reporting frameworks have begun to recognize the importance of prompt transparency at the clinical-trial level. CONSORT-AI[8] and SPIRIT-AI[9] address trials of AI interventions, while TRIPOD-LLM[10] emphasizes prompt disclosure and development reporting. However, current medical AI reporting standards do not provide a practical protocol for isolating prompt-component causality or detecting conditional effects of prompts on clinical outcomes. We developed PROMPT to fill this methodological gap.

PROMPT adapts the logic of pre-registered, randomized controlled trials. A matched-control prompt first removes the presumed active component while preserving or subtracting surrounding structure, task framing, and output format (analogous to placebo-controlled subtraction).[4-7] Its dismantling arms then isolate subcomponents including a factorial extension tests on whether components interact. Randomized prompt-stimulus trial assignment and blinded scoring complete the design. We provide PROMPT as a reusable protocol with prompt-arm templates, randomization procedures, decision rules, analysis templates, and example demonstration datasets.

We demonstrate the framework in two different settings. The first is a de novo tumbling-E optotype benchmark with objective orientation ground truth, and the second is a label-masked mammographic orientation task using CBIS-DDSM crops. The aim is not to validate a deployable mammography tool, but to show whether a mechanism identified in a synthetic benchmark transfers to a clinical-image feasibility task and whether the PROMPT framework itself generalizes across stimulus types. The two task design has complementary strength. The synthetic task provides controlled ground truth and complete factorial coverage at low cost; the clinical-image task tests whether the protocol reveals similar component structure in a setting with semantic complexity, inter-image variability, and real clinical content.

---

## Methods

### PROMPT Framework Design

PROMPT treats the prompt-stimulus trial as the experimental unit. Each stateless model call pairs one locked prompt arm with one locked stimulus, and every arm is crossed with every stimulus type (Figure 1). The full intervention prompt (Arm A) is compared with a matched-control prompt (Arm B), which functions as a prompt-level placebo control by removing the presumed active component while preserving the surrounding task framing, scaffold structure, output schema, and response format. A vehicle prompt (Arm C) provides an uninstrumented baseline. Dismantling and factorial arms (Arms D and E) remove or restore single subcomponents to test mechanism and interaction. In the present demonstrations the candidate active component is a task-specific decoding rule, which gives the model an explicit, mechanically verifiable mapping from glyph geometry (or anatomical landmarks, in the mammographic task) to one of four orientations; the second component is a prohibited-reasoning block, an instruction restricting the model from narrating or relying on free-form chain-of-thought before committing to an answer. Full verbatim text for every component and arm is provided in the Supplementary Appendix and the OSF archive.

Every arm is executed as stateless single-turn API calls; no conversational history or system prompt is carried between trials. Trial-prompt-stimulus assignment is randomized using a locked pseudorandom seed committed before execution. All prompt files are SHA-256 hashed; scoring of model outputs against the locked truth table is conducted blind to prompt-arm identity. This design means the experimental unit is the (prompt × stimulus) trial, not the session or the session sequence.

### Prompt Locking, Randomization, and Blinding

All experiments used Claude Sonnet 4.6 (claude-sonnet-4-6, Anthropic API). Trials were independent single-turn calls with no carried conversation history or system prompt. Prompt files, stimulus sets, truth keys, randomization seeds, hypotheses, and decision rules were committed before confirmatory testing. Prompt files were SHA-256 hashed; trial order used locked pseudorandom seeds (Experiment 1: seed 20260421; Experiment 2: seed 20260425); scoring was blinded to prompt arm. Testing was distributed across multiple days over approximately two weeks to reduce dependence on any single transient model service state. All five treatment files and the full reproducibility package, including SHA-256 hashes for independent verification, are deposited in the OSF archive: https://osf.io/7tuen/

### Experiment 1: Synthetic Tumbling-E Benchmark

Each tumbling-E chart contained 61 labeled E-shaped glyphs. Six tumbling-E probe configurations were analyzed: four uniform-orientation probes and two mixed-orientation probes. The main study included Arms A–C across seven probes, including an alphabetic Snellen probe. The Snellen probe was used as a specificity control. Across prompt arms, the model correctly rejected it as unsuitable for tumbling-E orientation detection, indicating that the prompts did not simply force orientation outputs on non-tumbling-E stimuli.

A dismantling mini-study tested Arms A–D on four uniform-orientation probes. A completed factorial extension tested Arms A–E across all probes. The primary outcome was glyph-level orientation accuracy, defined as the proportion of 61 glyphs per trial correctly classified to one of four directions (UP, RIGHT, DOWN, LEFT). Figure S3 shows an example probe and verbatim prompt excerpts for Arms A and B illustrating the matched-control structure.

**Experiment 2: Mammographic Feasibility Task**

Sixteen CBIS-DDSM[13] mammographic crops were label-masked and rotated to four orientations, producing 64 stimuli. Ground truth was defined by an investigator-annotated chest-wall-to-nipple vector. The confirmatory run tested Arms A-D; a pre-registered Arm E extension completed the factorial comparison. The task was intended as a feasibility demonstration with clinical images, not as a diagnostic validation study.

As described in the primary data publication,[13] the original de-identified and publicly available CBIS-DDSM dataset was collected under IRB approval with patient consent waiver. The imaging assets are hosted via the Cancer Imaging Archive (TCIA)[14]. Prospective specification of our analytic pipelines follows the broader methodology of the clinical preregistration revolution[15].

**Statistical Analysis**

Experiment 1 primary analysis used a linear mixed model (LMM) with arm as fixed effect and probe as random intercept (restricted maximum likelihood; one-sided z-test $P < 0.05$). The probe-level intraclass correlation coefficient was 0.337, confirming that 34% of total glyph-accuracy variance is attributable to probe identity and justifying the probe as a blocking factor. As sensitivity checks, the paired t-test using 180 probe-matched pairs yielded the same A−B difference, with $d_z$=1.12, $t(179)$=15.0, and $P=9.6\times10^{-34}$; the independent t-test yielded $d$=1.60 and $P=8.1\times10^{-41}$. These analyses supported the primary conclusion, although the effect-size estimates differed. Dismantling and factorial contrasts used pre-specified probe-level decision rules, Bonferroni correction across P1, P3, and P4 where applicable, and an A−E equivalence criterion of less than 10 percentage points.

Experiment 2 primary analysis (H-mammo-1) tested whether A−B  +15 percentage points with bootstrap 95% confidence interval lower bound > 0 (5,000 iterations; locked seed 20260425; unpaired trial-level resampling). A post-hoc LMM sensitivity analysis with stimulus as random intercept (ICC = 0.774) was also conducted. Arm B error-pattern uniformity was tested by chi-square goodness-of-fit; Cramér's V quantified effect size. Wilson 95% confidence intervals were computed for binomial proportions.

Invalid and unparsable model outputs were excluded from accuracy denominators; parse rates are reported per arm. In Experiment 2, excluded unparsable outputs were 0 of 64 trials in Arm A, 0 of 64 in Arm B, 5 of 64 in Arm C, 2 of 64 in Arm D, and 0 of 64 in Arm E. Per-orientation mammography findings were treated as hypothesis-generating because each orientation cell contained 16 trials.

**Reproducibility Materials**

The protocol package include prompt arms, locked seeds, trial logs, truth tables, response files, analysis code, figure-generation files, and SHA-256 hashes. This reporting structure follows the resource-oriented pattern of Datasets, Benchmarks, and Protocols articles. The reusable contribution is PROMPT itself, and the two experiments serve as worked demonstrations of the protocol. A step-by-step PROMPT workflow with component-definition guidance, prompt-arm templates, pre-registration checklist, and analysis code is provided in the Supplementary Appendix and in the OSF archive. A minimal PROMPT run on a new task requires only a lockable stimulus set, a candidate active component, and approximately 150 API trials (five arms  30 trials per stimulus type), costing approximately 15 USD for a synthetic benchmark.

---

## Results

**PROMPT generated interpretable component-level findings**

Figure 1 summarizes the PROMPT workflow used to generate component-level inference across both demonstrations. Across the two tasks, PROMPT distinguished active, inactive, harmful, and task-dependent prompt components rather than only whole-prompt performance differences. The same canonical-direction failure mode appeared in both tasks, suggesting that the error phenotype reflected prompt structure and model priors rather than a stimulus-specific artifact. Table 1 summarizes the prompt arms, pre-registered hypotheses, decision rules, and key outcomes.

**Synthetic benchmark identified the active component**

In the tumbling-E main study, Arm A achieved 98.6% glyph-level accuracy, compared with 50.1% for Arm B and 74.8% for Arm C (Figure 2; Table 1). The pre-registered primary contrast was supported: A−B = +48.5 percentage points (95% CI, +43.2 to +53.7; $z=18.1$; $P=1.9\times10^{-73}$; linear mixed model with probe as random intercept; Table 1; Table S2). Arm B performed below Arm C (B−C = −24.7 percentage points; $z=-8.6$; $P<10^{-17}$), showing that scaffolded prompt structure without the decoding rule was actively harmful rather than neutral (Table 1; Table S2). Arm B accuracy was strongly bimodal across probes rather than centered on its mean (Figure 2A): some probes were classified near ceiling while others collapsed toward the canonical-RIGHT prior, so the 50.1% mean reflects a mixture of near-correct and near-collapsed probes rather than uniformly intermediate performance. Mechanistically, this is consistent with the matched-control scaffold suppressing the heuristic fallback that the uninstrumented vehicle prompt leaves available, rather than with the scaffold introducing errors de novo.

This finding is clinically important because Arm B retained the structural elements of the full intervention (i.e. task framing, prohibited-reasoning block, and output schema) but lacked the task-specific decoding rule. Without a matched-control arm, a developer comparing only complete prompt variants would not detect that the scaffolded prompt structure itself was harmful in the absence of the rule.

**Error phenotype was systematic, not random**

Arm B errors were highly structured rather than diffusely distributed (Figure 3). Confusion-matrix analysis showed canonical-RIGHT collapse and DOWN-to-UP mirror confusion in no-rule arms (Figure 3). In Arm B, 63% of incorrect glyph responses predicted RIGHT regardless of truth orientation; Arm D, which removed both the decoding rule and the prohibited-reasoning block, showed a similar no-rule error phenotype (Figure 3; Table S2). This pattern supports interpretation of the canonical-RIGHT collapse as a systematic directional bias reflecting the model's prior toward the standard orientation of the letter E.

A post-hoc per-glyph analysis showed that Arm A residual errors were concentrated on Row 11, the smallest 20/10 acuity line: 120 of 122 wrong-glyph responses occurred there (Figure S1; Table S2). Rows 1–10 were essentially error-free, with only two wrong-glyph responses among 9360 scored glyphs, corresponding to an error rate of 0.02% (Figure S1; Table S2). This exploratory result supports a resolution-limit interpretation for the full-prompt arm. Separately, the canonical-RIGHT collapse represented a systematic directional-bias phenotype, a safety-relevant failure mode distinct from diffuse inaccuracy, which PROMPT detected reproducibly across the synthetic and mammographic tasks.

**Factorial testing showed rule independence in the synthetic task**

The completed 2×2 factorial showed that Arm E, the rule-only arm, tracked Arm A across all six tumbling-E probes (Figure 2; Table S3). The maximum A−E difference was 2.3 percentage points, within the pre-specified equivalence criterion (Table 1; Table S3). E−D was +85.2 percentage points on LEFT, +46.7 on UP, and +90.0 on DOWN; all three pre-specified contrasts cleared the Bonferroni-corrected threshold (Table S3). Thus, in the synthetic benchmark, the decoding rule was the sole measurable active component.

The pooled LMM across the three pre-specified probes confirmed the direction of the factorial result: E−D = +74.0 percentage points; $z=16.9$; $P=6.7\times10^{-64}$ (Table S3). The prohibited-reasoning block added nothing when the rule was present, with A−E ≤2.3 percentage points across all probes (Figure 2; Table S3). In the dismantling study, removing the prohibited-reasoning block from Arm B was harmful on P3 and beneficial on P4, confirming that its effect depended on both orientation and rule presence (Table S3).

**Clinical-image feasibility showed attenuation and task dependence**

In the mammographic orientation task, Arm A achieved 73.4% accuracy and Arm B achieved 59.4% accuracy (Figure 4; Table S4). The A−B difference was +14.1 percentage points (95% CI, −3.1 to +29.7), below the pre-specified threshold; therefore, H-mammo-1 was not supported (Table 1; Table S4). However, Arm B errors were non-uniform and concentrated on canonical RIGHT predictions ($\chi^2$=19.23, df=3, $P<0.001$), supporting H-mammo-2 and showing that the canonical-RIGHT error phenotype identified in Experiment 1 recurred on clinical mammographic stimuli (Figure 4; Table S4).

The Arm E extension showed that removing the prohibited-reasoning block reduced accuracy on DOWN-orientation images by 25.0 percentage points, from 62.5% in Arm A to 37.5% in Arm E, but had no effect on UP, RIGHT, or LEFT orientations (Figure 4; Figure S2; Table S4). This 25.0-percentage-point difference rests on a single 16-trial cell, was not adjusted for the four orientations examined, and carries a wide confidence interval (Wilson 95% CI for the difference spans zero); it should be read as a hypothesis to be tested in a powered replication rather than as an established effect. With that caveat, the same block that added no measurable benefit in the synthetic task was numerically associated with resistance to canonical-direction collapse specifically on the orientation most conflicting with the model's canonical bias. Because each orientation cell contained 16 trials, these per-orientation estimates are feasibility-level and hypothesis-generating.

A post-hoc linear mixed model with stimulus as random intercept yielded A−B = +14.1 percentage points (95% CI, +3.5 to +24.6; z=2.61; P=0.009), confirming a positive effect while remaining below the pre-specified +15 percentage-point threshold (Table S4). Thus, under both the pre-specified bootstrap analysis and the post-hoc LMM sensitivity analysis, the mammographic task showed an attenuated rule benefit rather than confirmatory support for H-mammo-1.

---

### Table 1. PROMPT protocol elements, decision rules, and key results.

| Protocol element | Implementation in PROMPT | Key result or decision |
|---|---|---|
| **Matched-control comparison** | Arm A: full intervention; Arm B: matched control without decoding rule | Tumbling-E H1 supported: A-B = +48.5 pp; mammography H-mammo-1 not supported: A-B = +14.1 pp |
| **Vehicle baseline** | Arm C: minimal uninstrumented prompt | Arm B performed below Arm C in tumbling-E, showing harmful scaffold-without-rule behavior |
| **Dismantling arm** | Arm D: no rule, no prohibited-reasoning block | Direction-dependent effects showed that the prohibited-reasoning block did not have a single global effect |
| **Rule-only factorial arm** | Arm E: decoding rule without prohibited-reasoning block | In tumbling-E, Arm E tracked Arm A; maximum A-E difference = 2.3 pp |
| **Clinical-image transfer** | Same arm structure applied to rotated mammographic crops | Canonical-RIGHT collapse reproduced; accuracy rescue attenuated; DOWN-specific Arm A vs Arm E signal was exploratory |
| **Reproducibility package** | Locked prompts, seeds, stimuli, truth tables, logs, analysis files, hashes | Supports reuse of PROMPT as a protocol rather than only a single experimental result |

# Discussion

PROMPT introduces a pre-registered, randomized framework for evaluating prompt engineering as a component-level intervention in medical AI. Existing reporting guidelines, including CONSORT-AI, SPIRIT-AI, and TRIPOD-LLM, have strengthened transparency by requiring clearer reporting of AI interventions, protocols, and prompts.[8–10] However, these guidelines do not provide an experimental design for isolating and estimating the causal contribution of individual prompt components. PROMPT addresses this gap by moving prompt development beyond informal iteration and whole-prompt comparison toward prospective, controlled, component-level evaluation.

The core innovation is the structured comparison of prompt arms. The matched-control arm functions as a prompt-level placebo control: it removes the presumed active component while preserving surrounding task framing, scaffolding, output schema, and response format. Dismantling and factorial arms then test whether subcomponents act independently or interact. In the synthetic benchmark, this design showed that scaffolding without the task-specific decoding rule was actively harmful compared with the vehicle prompt. A conventional A/B comparison could have shown that the full prompt outperformed the matched control, but it would not have identified that the scaffolded structure became harmful when separated from the rule. This finding supports the need to evaluate prompts as structured interventions rather than as indivisible text blocks.

PROMPT also showed that prompt evaluation should assess error structure, not only average accuracy. In rule-absent arms, the model showed a canonical-RIGHT error phenotype, consistent with a prior toward the standard presentation of the letter E. The same collapse appeared when both the decoding rule and prohibited-reasoning block were absent, suggesting that the bias was not created by the prohibited-reasoning block. Instead, that block appeared to amplify or modulate a pre-existing prior depending on task and orientation. This distinction is safety-relevant: systematic errors can recur predictably in specific subgroups, orientations, or clinical contexts even when aggregate accuracy appears acceptable.

The tumbling-E benchmark was useful because it made this mechanism legible. The task is deliberately geometric, with no clinical context or semantic cues from which orientation can be inferred. Without the decoding rule, the model can fall back on a learned presentation prior; with the rule, it has a mechanically verifiable mapping from glyph structure to direction. The Row 11 finding adds a complementary mechanism. Even when the rule was present, residual errors concentrated on the smallest 20/10 acuity line, where visual resolution limited reliable rule application. Thus, the synthetic benchmark separated a prompt-boundary failure, in which the rule was absent, from a stimulus-boundary failure, in which the rule was present but the glyph was too small to resolve.

The mammographic feasibility task tested whether these mechanisms transferred to semantically complex clinical images. The canonical-RIGHT phenotype recurred, suggesting that the error pattern was not merely an artifact of the tumbling-E stimulus set. However, the rule's benefit was attenuated. This is expected because the tumbling-E rule is geometric and mechanically verifiable, whereas the mammographic rule requires identifying anatomical landmarks, including the chest wall and nipple, whose appearance varies with rotation, tissue density, and source-image characteristics. The two-task design therefore revealed both transfer and boundary conditions: the error phenotype transferred, but the rule's ability to rescue performance weakened as the task became clinically and semantically more complex.

The prohibited-reasoning block illustrates why prompt effects should not be treated as globally beneficial or harmful. In the synthetic task, the block was inactive when the decoding rule was present. In the mammographic task, it appeared to contribute selectively on DOWN-oriented images, where conflict with the model's canonical orientation prior was strongest. This per-orientation result is feasibility-level evidence, but it illustrates a general safety principle: prompt components may be redundant in one task, harmful in another, and beneficial in a narrow subgroup. Such conditional effects are difficult to detect with conventional prompt-engineering evaluations, which typically compare complete prompts rather than individual components.[5,6]

PROMPT was designed to make prompt evaluation auditable and reproducible. The framework requires prospective documentation of prompt arms, hypotheses, decision rules, stimuli, truth tables, randomization seeds, and analysis plans. In this study, the OSF archive includes treatment files, locked stimuli, truth tables, trial logs, response files, analysis scripts, figure-generation files, app-runner materials, and SHA-256 hashes. This structure allows independent verification of what was tested, when it was locked, how trials were randomized, and how outputs were scored. This emphasis on prospective specification and reproducibility aligns with broader efforts to strengthen preregistered and early-stage clinical AI evaluation.[15,16]

The framework is also practical. A minimal five-arm synthetic benchmark can be conducted at modest API cost, allowing prompt-component testing early in development rather than only after a system has reached clinical validation. PROMPT should therefore be viewed as a complement to, not a replacement for, downstream clinical validation. Existing reporting standards help ensure that AI interventions and prompts are disclosed transparently; PROMPT adds an experimental design for determining which prompt components shape model behavior and under what conditions. By making prompt engineering testable, verifiable, and reproducible, PROMPT offers a practical step toward safer clinical AI development.

## Limitations

This study has several limitations. Most importantly, the effective sample size of the mammographic analyses is modest: because the 16 source images each contributed four rotated stimuli and stimulus identity accounted for a large share of the variance (post-hoc stimulus-level ICC = 0.774), the effective n is much closer to the 16 source images than to the 64 stimuli. Mammographic ground truth was annotated by a single investigator, per-orientation mammography cells were small, and each source image contributed multiple rotated stimuli. The mammographic results should therefore be interpreted as feasibility-level evidence, not clinical validation. The study also used one foundation model and did not include a zero-prompt baseline to directly measure the model's orientation prior. The empirical findings are model- and task-specific, although the PROMPT protocol itself is model-agnostic and can be applied to other API-accessible systems.

## Conclusions

When prompts shape clinical AI behavior, they should be evaluated as interventions: pre-specified, controlled, and transparent. PROMPT shows that component-level prompt testing is feasible at low cost and reveals findings not available from whole-prompt comparisons. The present demonstrations address spatial-orientation tasks; the protocol is in principle applicable to other clinical AI tasks in which prompt components may alter behaviour, such as radiology interpretation, pathology review, triage, documentation, and decision support, though extension to those settings remains to be demonstrated.

---

## Acknowledgments

The author acknowledges methodological feedback from large language model reviewers used as thinking partners during protocol development and manuscript drafting. AI methodological assistance is disclosed per emerging community norms; no AI system was a co-author or contributed substantively to the scientific content beyond methodological review and drafting support.

## Data and Code Availability

All prompt arms, locked seeds, stimuli, truth tables, response logs, analysis scripts, figure-generation files, and SHA-256 hashes are deposited in the OSF archive for this study. The archive is intended to allow independent reproduction of all main figures and tables and reuse of the PROMPT protocol templates.

---

## Figures

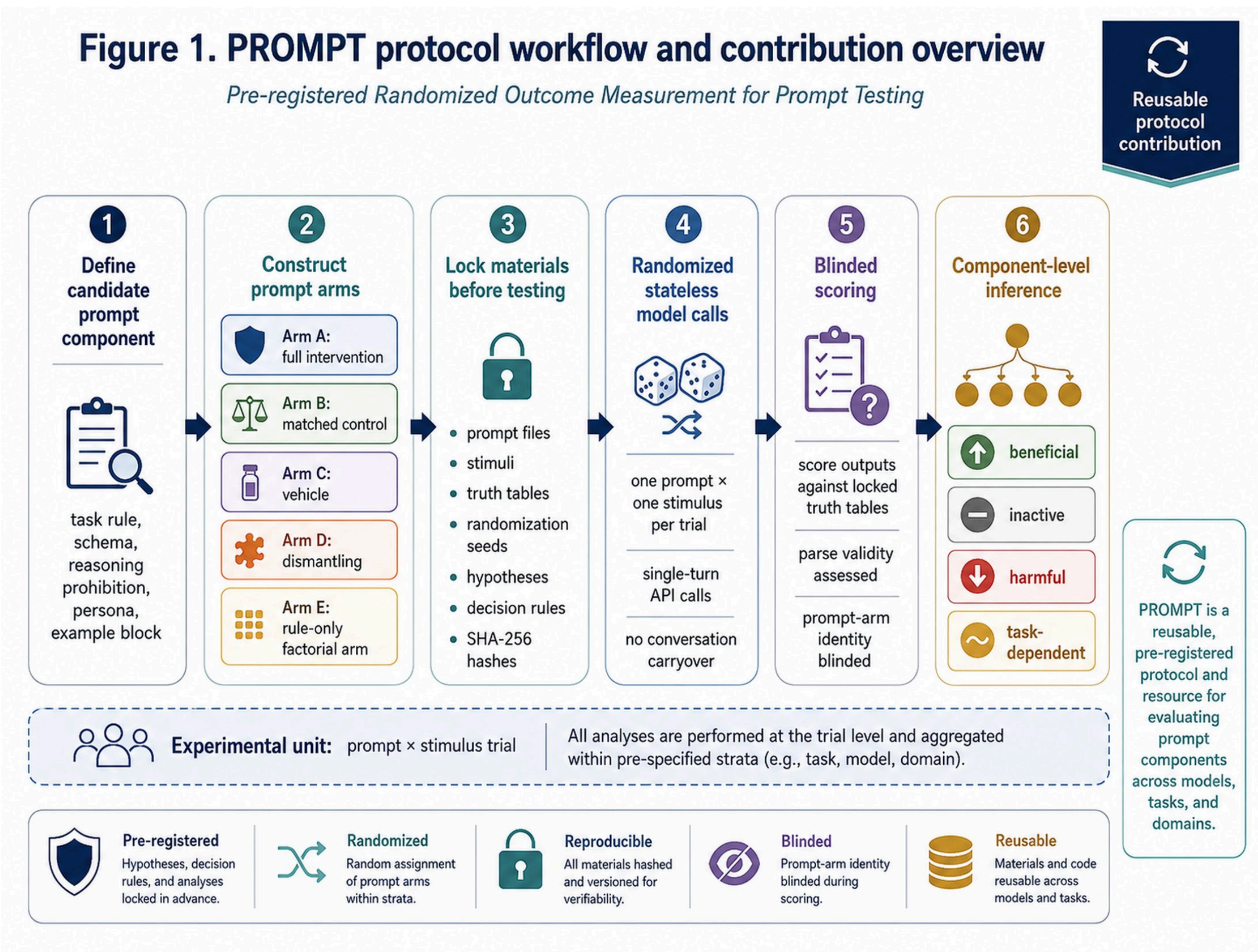

**Figure 1. PROMPT protocol workflow and contribution overview.** The reusable protocol begins with component definition and prompt-arm construction, then locks prompts, stimuli, truth tables, randomization seeds, hypotheses, and decision rules before randomized stateless model calls and blinded scoring. The output is component-level inference about active, inactive, harmful, and task-dependent prompt components.

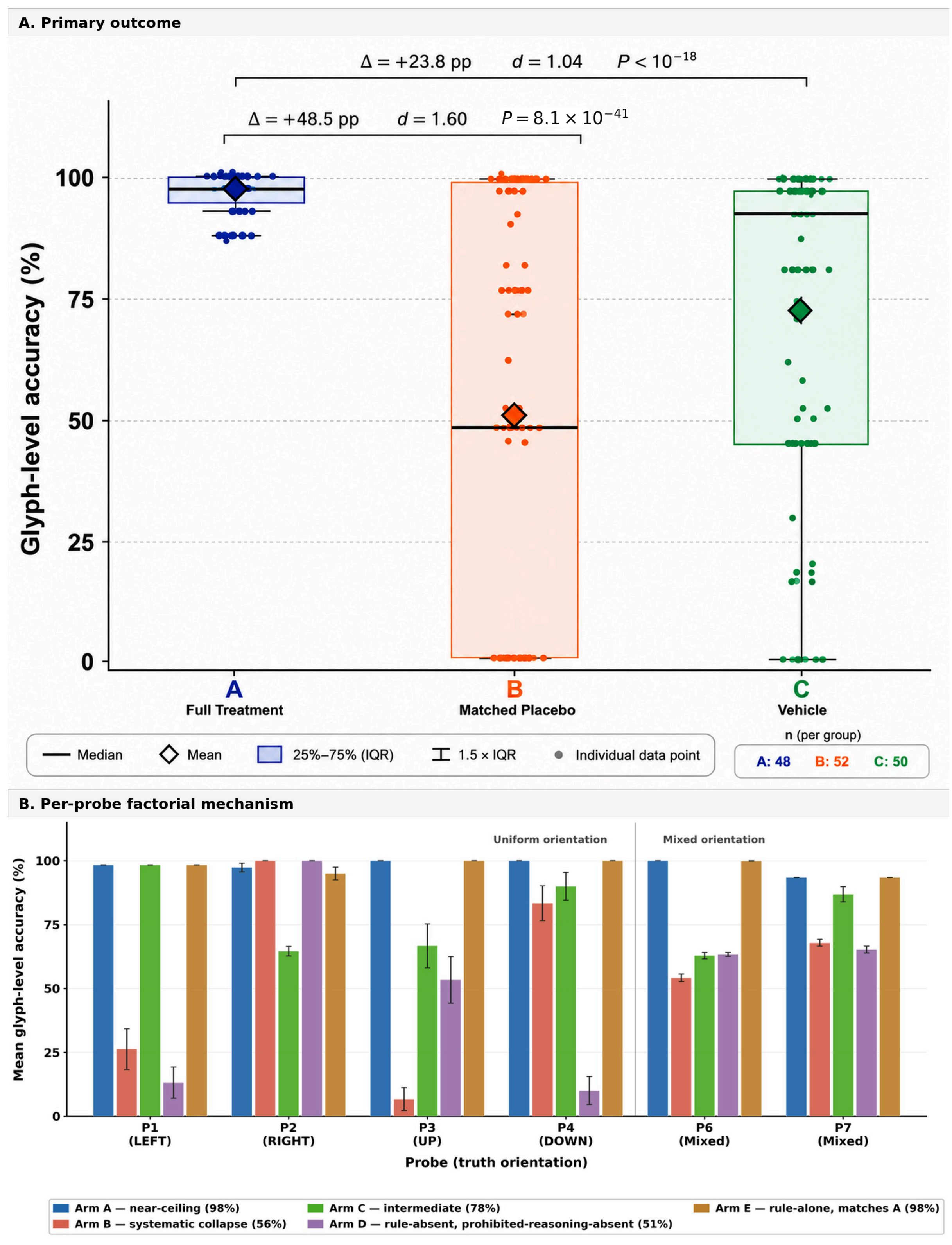

**Figure 2. Experiment 1 synthetic benchmark outcomes and factorial mechanism.** Panel A shows trial-level glyph accuracy in the tumbling-E main study. Panel B shows mean glyph-level accuracy per arm-probe cell across the completed five-arm factorial. Arm E tracks Arm A across

all six tumbling-E probes, supporting the conclusion that the decoding rule is the sole measurable active component in the synthetic task.

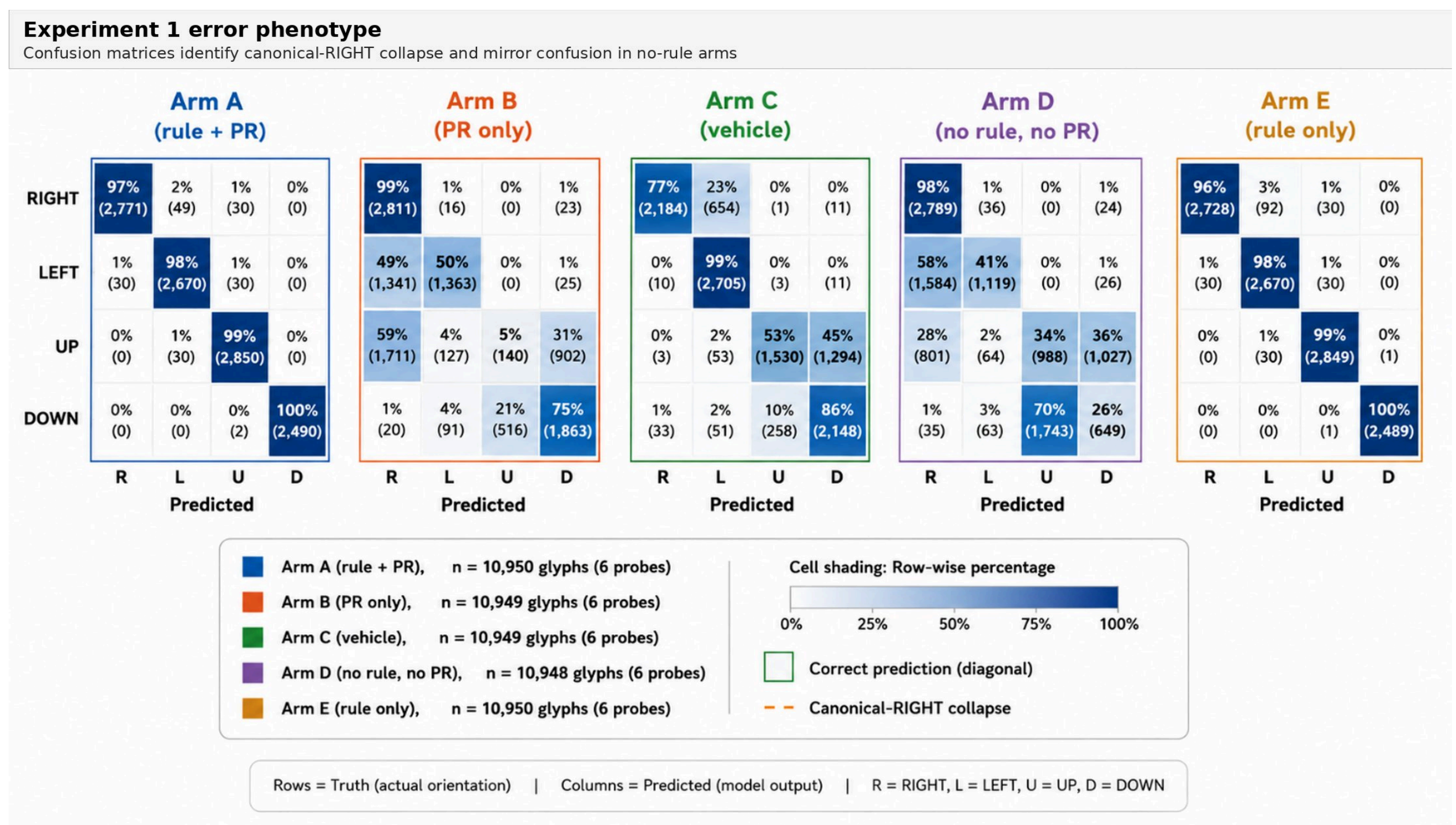

**Figure 3. Experiment 1 error phenotype in the tumbling-E task.** Per-arm confusion matrices show near-diagonal performance in rule-containing arms and systematic canonical-RIGHT collapse and DOWN-to-UP mirror confusion in no-rule arms. This figure emphasizes that matched-control failure was structured rather than random.

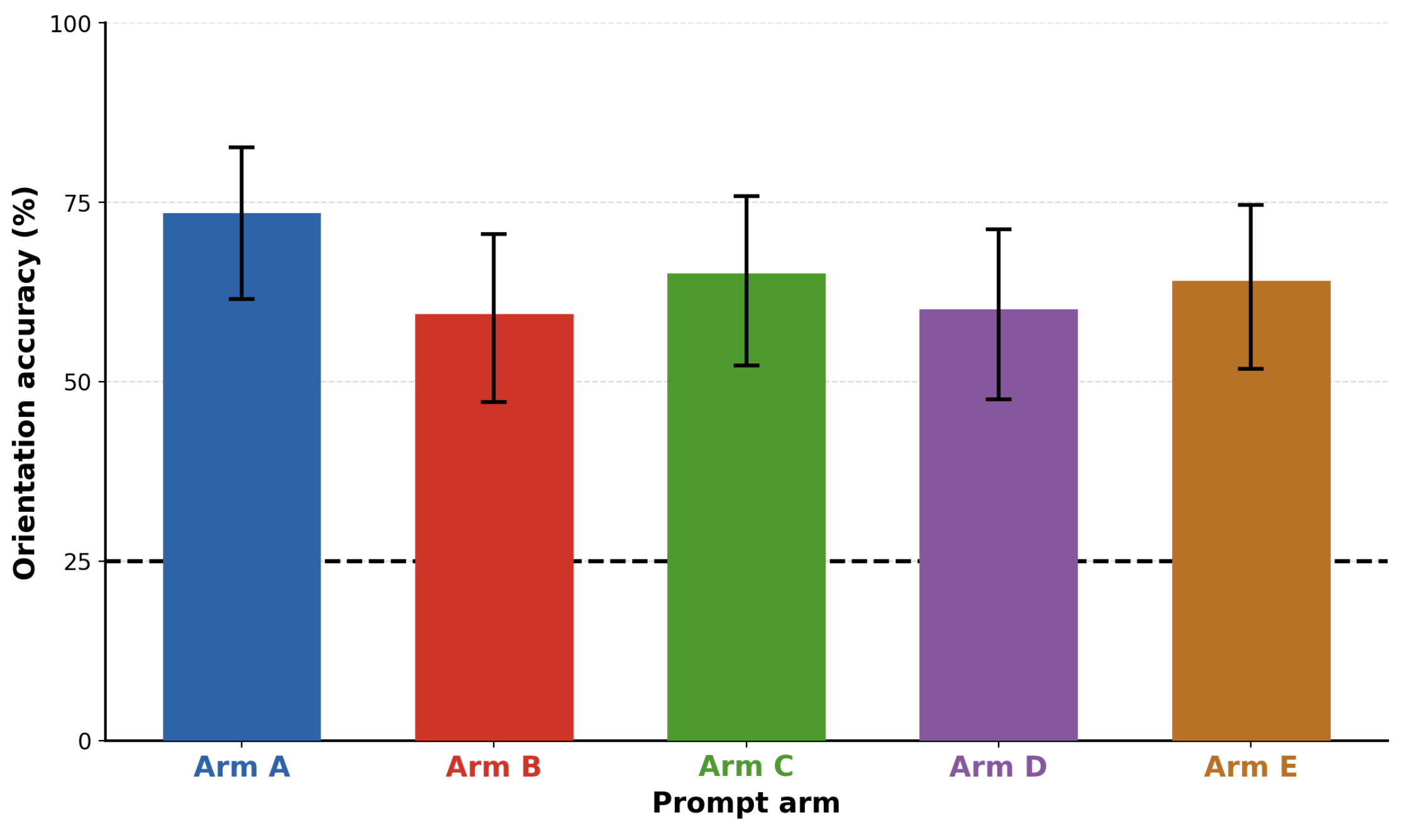

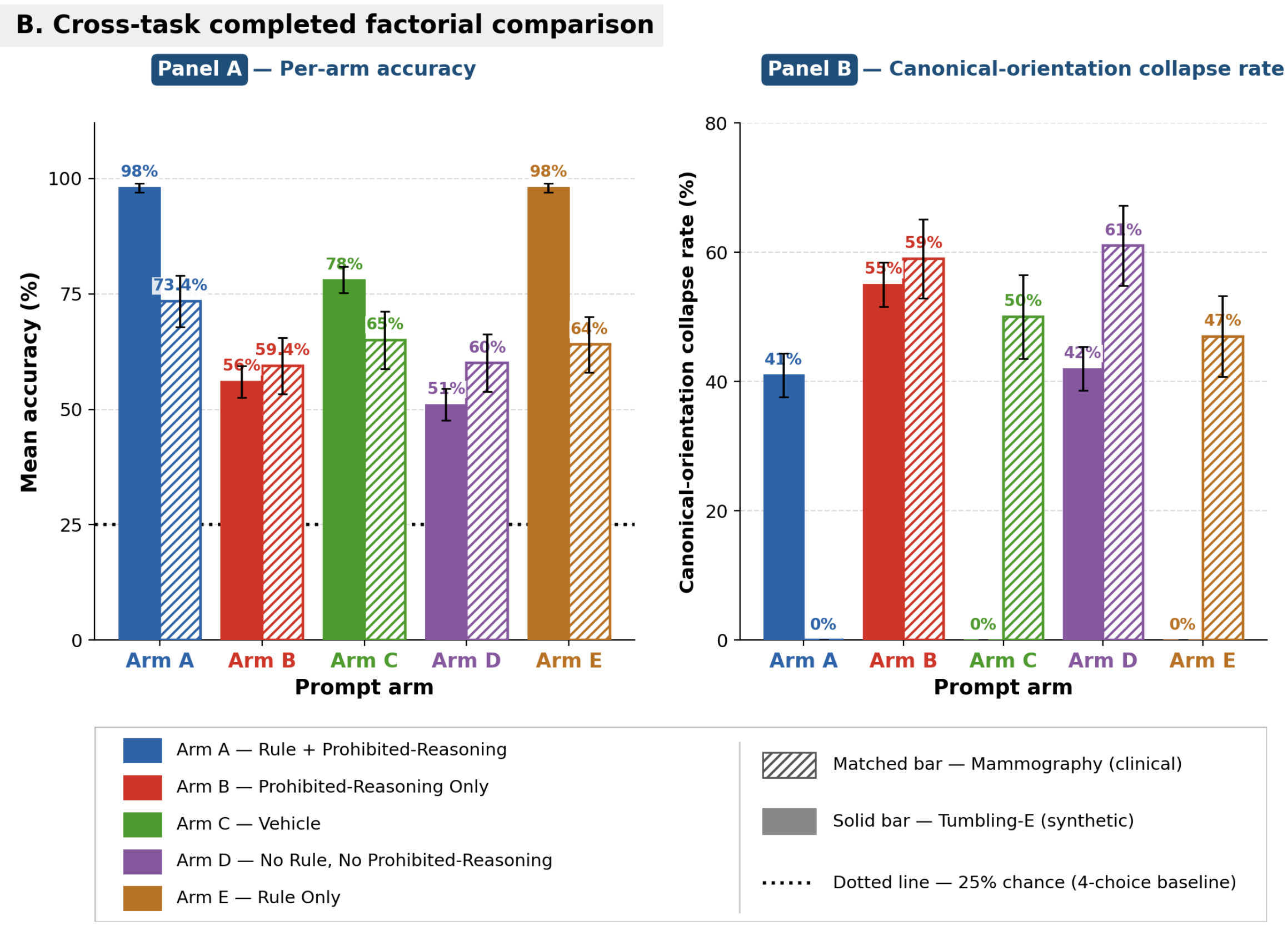

**Figure 4. Mammographic feasibility task and cross-task transfer.** Panel A shows per-arm mammographic orientation accuracy. Panel B compares completed factorial behavior across the synthetic and mammographic tasks. The decoding rule transferred as an error-pattern signal but with attenuated accuracy rescue in clinical images; the prohibited-reasoning block showed a task-dependent contribution not observed in the synthetic task.

---